# A GPS spoofing detection and classification correlator-based technique using the LASSO

Erick Schmidt, *Student Member, IEEE*, Nikolaos Gatsis, *Member, IEEE,* and David Akopian, *Senior Member, IEEE*

*Abstract*—This work proposes a global navigation satellite system (GNSS) spoofing detection and classification technique for single antenna receivers. We formulate an optimization problem at the baseband correlator domain by using the Least Absolute Shrinkage and Selection Operator (LASSO). We model correlator tap outputs of the received signal to form a dictionary of triangle-shaped functions and leverage sparse signal processing to choose a decomposition of shifted matching triangles from said dictionary. The optimal solution of this minimization problem discriminates the presence of a potential spoofing attack peak by observing a decomposition of two different code-phase values (authentic and spoofed) in a sparse vector output. We use a threshold to mitigate false alarms. Furthermore, we present a variation of the minimization problem by enhancing the dictionary to a higher-resolution of shifted triangles. The proposed technique can be implemented as an advanced fine-acquisition monitoring tool to aid in the tracking loops for spoofing mitigation. In our experiments, we are able to distinguish authentic and spoofer peaks from synthetic data simulations and from a real dataset, namely, the Texas Spoofing Test Battery (TEXBAT). The proposed method achieves 0.3% detection error rate (DER) for a spoofer attack in nominal signal-to-noise ratio (SNR) conditions for an authentic-over-spoofer power of 3 dB.

## I. Introduction

GLOBAL navigation satellite systems (GNSS) such as the Global Positioning System (GPS) [1] provide crucial positioning and timing for applications in the civil, commercial, and military domains. Recently, GNSS receivers have grown in popularity due to their low costs and broad applications. Instances of GNSS uses can be seen in financial transactions, phase measurement units (PMUs) in power grids, and emergency services [2].

The open-access aspect of the GPS coarse acquisition (C/A) codes exposes the system to potential malicious attacks to position and timing-dependent applications. Such unintentional or intentional attempts are categorized as *jamming* and *spoofing*. While *jamming* attempts to disrupt or degrade GPS channels by signal blocking or overpowering, a smarter and more hazardous *spoofing* attack can imitate GPS signals aiming to mislead the target receiver and infringe flawed position and timing resolutions. The vulnerability to GNSS spoofing is an active research area due to its impact in critical and ever-growing GNSS-dependent applications [2].

Once the target receiver is deceived into locking to counterfeit signals, the typical spoofing attack shifts the authentic code and carrier phases to alter the position, velocity, and timing (PVT) solutions. Typically, commercial off-the-shelf (COTS) receivers lack ability to detect spoofing attacks, as has been proven in [3]. Additionally, recent software-defined radio (SDR) platforms have demonstrated fast-prototyping for spoofing attack implementation and mitigation techniques that otherwise commercial receivers lack [4]. Literature has categorized the type of spoofing attacks into simplistic, intermediate, and advanced [5] based on the complexity of the spoofing device, with intermediate spoofing being the most cost-effective in terms of implementation.

### A. Multipath considerations

Often, spoofing attacks can manifest as multipath (MP) [6], [7]. In fact, considerable research addresses the discrimination between spoofing and MP [7], [8], [9]. However, there are four overall main differences: (1) the delay profile of the authentic and spoofed signal combined appears to be sparse per channel, as opposed to MP signals which appear as a cluster of reflected signals with various delays referred to as delay profile [10]; (2) the spoofing attack occurs on all channels concurrently; (3) the incurred attack delays appear similar on all channels; and (4) such attacks can overall incur significantly more damage to the PVT solution, e.g., cause the user position and time estimates to deviate more substantially when compared to MP. Therefore, this work focuses on a detection and classification technique particularly for spoofing attacks. In the next subsection, we provide a literature review on anti-spoofing techniques including the most relevant MP techniques for the sake of categorization. Further, a qualitative comparison of state-of-the-art spoofing and MP countermeasures in the baseband domain is provided in Section VI-B.

This manuscript was first submitted to the IEEE Transactions on Aerospace and Electronic Systems on August 23, 2019. It was resubmitted on February 5, 2020 for MAJOR CHANGES.

This work was supported by the National Science Foundation under Grant ECCS-1719043.

E. Schmidt, N. Gatsis and D. Akopian are with the Department of Electrical and Computer Engineering, The University of Texas at San Antonio, San Antonio, TX, 78249 USA (e-mail: erickschmidtt@gmail.com; nikolaos.gatsis@utsa.edu; david.akopian@utsa.edu).

*B. Spoofing countermeasures*

In recent literature, GNSS spoofing countermeasures have been categorized based on numerous aspects of proposed techniques and receiver implementation domain. In the following, we categorize spoofing countermeasure *techniques* and their *extent* based on [5], [11], [12], and [13]. Fig. 1 shows a categorization map where an asterisk narrows down the discussion in this work.

The countermeasure techniques according to Fig. 1 fall into four main categories [5]: (1) single-antenna advanced signal processing-based methods, (2) encryption-based methods, (3) drift monitoring methods, and (4) signal-geometry-based or multi-antenna methods. Signal processing-based methods rely on receiver tracking loops [4], correlator outputs [6], [9], [10], automatic gain control (AGC) power monitoring [7], and vector tracking loops (VTL) [12], [14]. There are encryption-based signal authentication methods that are yet to be implemented in civilian GNSS signals [15], [16]. Drift monitoring methods identify unexpected variations in the positioning or timing solutions [17], [18], [19]. Finally, signal-geometry-based or multi-antenna methods rely on estimating the angle-of-arrival or spatial vector between authentic and counterfeit signals [20], [21], [22], [23], [24]. Furthermore, authors in [11] categorize spoofing countermeasures into baseband domain—related to techniques pertaining to signal acquisition and tracking in the physical layer [4], [6], [9], [10]—and navigation domain such as receiver autonomous integrity measurement (RAIM) [25]. The baseband domain is further sub-categorized into pre-correlator [26], [27], correlator [28], and post-correlator [12], [29] domains.

In terms of spoofing *countermeasure extent*, the techniques can be classified into three independent categories [30]: (a) *detection*, which can be also seen as a binary decision monitor usually based on scalar-valued output metrics [9]; (b) *classification*, which discerns patterns in the received signal based on the nature of the technique, e.g., a MP delay profile [10], auxiliary peak tracking [31], or chip-level MP delays [6]; and (c) *mitigation*, which provides correction or rejection of the attack [7]. Also, these categories are considered independent such that for example, one countermeasure technique may have detection, detection and mitigation, or all three.

*C. Contributions of This Paper*

This paper addresses intermediate spoofing attacks based on a single-receiver single-antenna advanced signal-processing technique with a *detection* and *classification* extent. The proposed method falls into the *baseband correlator* domain (see Fig. 1). This domain is critical because it precedes navigation, where the damage to the PVT solution is by that time rendered. GNSS signals are commonly processed using a correlation-based synchronization of received signals with locally generated replicas of expected signal patterns. In particular, an ideal correlation profile of a GPS C/A signal resembles a triangle function, where the triangle elements correspond to the correlations of the received signal with replica fragments generated with various time delays. The triangle peak corresponds to the correlation with the aligned replica.

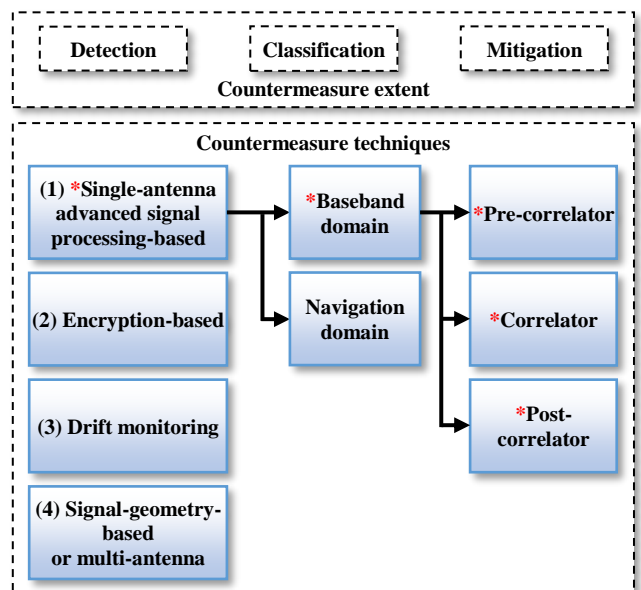

Fig. 1. Spoofing countermeasure categorization map and potential countermeasure extent.

Spoofing signals distort the triangle profile and complicate the synchronization process, as the correlation profile becomes a superposition of several such triangles of unknown intensity. In addition, such distortions can be mixed with residual uncompensated sinusoidal modulations due to Doppler effects.

This paper develops an automatic method for triangle-based decomposition of the correlation profiles and extraction of contributing individual components, resulting from both desired and spoofing signals. The proposed decomposition exploits an optimization problem modeling the Least Absolute Shrinkage and Selection Operator (LASSO) [32]. Then, the decomposition helps to discriminate desired and spoofing components via a sparse output. We characterize the correlation profile of the received signal using a dictionary of shifted triangle shapes and a sparse vector to select potential shifted triangles from said dictionary. The optimal solution of this minimization technique discerns the presence of a spoofing attack by observing two different code-phase values, i.e., authentic and spoofed peaks, in the sparse vector. In addition, we use a threshold to mitigate false alarms.

Moreover, we present a variation of the minimization problem by enhancing the dictionary to a *higher resolution* of shifted triangles. Specifically, the higher resolution aspect improves the detection capability (sensitivity) such that a peak appearing between two discrete code-phase sampling points is still detected, while the correlator configuration remains unchanged. Finally, three concepts are presented to validate the techniques via Monte-Carlo simulations: (1) peak sensitivity response (PSR) curves, for sensitivity analysis; (2) peak detection error rate (DER) curves, for performance analysis; and (3) probability of false alarm (PFA).

The signal processing of the proposed technique relies on discerning two steps occurring in the tracking loops: correlation and integration or so-called *integrate-and-dump* filter [1], and tracking loop discriminators and feedback filters. We specifically analyze the correlator taps after carrier wipe-off and before entering the discriminators and feedback loop filters. The method is proposed to detect a spoofing event and discriminate when two peaks are present. Additionally, the technique is not suggested as a replacement module for conventional GPS receivers, rather as a baseband advanced fine-acquisition monitoring tool that can be deployed based on alarm-threshold strategies, or on scheduled or other arbitrary times. Further, by discerning between authentic and counterfeit peaks, the tracking loops can intelligently decide to follow the authentic peak without additional complex modifications. As long as the COTS receiver provides correlator tap outputs, the proposed monitoring tool can potentially be coupled with additional algorithms such as auxiliary peak tracking [30], [31], or advanced navigation-level spoofer-detectors [17], [18], [19]. To the best of our knowledge, we are the first to contribute on the following specific components:

1) We specifically model spoofing as a characteristic sparse event, i.e., spoofing peaks appear discretely, and thus can be addressed via sparse techniques.
2) The LASSO is used as an optimization technique for automatic peak discrimination.
3) A high-resolution aspect is introduced to the discrimination process, further discussed in Section IV.
4) A *multi-LASSO* optimization problem enhances the discrimination of spoofer peaks that appear between two discrete code-phase sampling points.

Without losing generality, the GPS C/A code signal is used throughout this paper, but the proposed technique can be extended to other GNSS signals.

The paper is organized as follows. Section II presents the signal model and spoofer overview. Section III formulates the problem and presents the LASSO based method. Section IV expands to another variation based on LASSO and formulates the PSR concept. Section V presents the testing methodology and Monte-Carlo simulations, and presents results for synthetic data and a real dataset. Section VI discusses related work. Finally, Sections VII and VIII respectively present concluding remarks and future work.

## II. SIGNAL MODEL AND SPOOFER OVERVIEW

The overall function of a GPS receiver is to maintain continuous synchronization with visible satellite signals for range measurements, ephemeris data extraction, and PVT estimation. This synchronization is achieved in two steps: (coarse) *acquisition* to find visible satellite signals and (fine) *tracking* for regular operation [33].

### A. Authentic signal model

Conventional GPS receivers use tracking loops for joint fine-tuning of the incoming signal to residual Doppler carrier frequency and phase offsets, and spreading code alignment. The phase lock loop (PLL) tracks carrier-phase alignments, and the delay lock loop (DLL) tracks code-phase alignments. Both loops achieve this by generating local carrier and code replicas, respectively. Discriminators and filters for both the PLL and DLL are used afterwards as feedback loops. An initial estimation of a number of received spreading code chips against the locally generated code replica is commonly called a *code-phase*. A set of correlators in the DLL compare several phase-shifted copies of the local code replica with the incoming signal for *code-phase* adjustments. COTS receivers typically employ three shifts to find the peak of the correlators, namely early, prompt, and late (EPL) correlators, however, advanced receivers with higher resolution in code-phase tracking loops are reported with hundred or more correlators [34]. The correlator spacing is typically within a 1-chip period. This allows *code-phase* synchronization with at least one replica with sub-chip accuracy [1], [33].

A GPS signal seen at a single-antenna receiver front-end is composed of an ensemble of satellite signals (channels) and their corresponding interference plus noise. Without loss of generality, the complex-valued baseband received signal for a single GPS channel, $l$, can be modeled after RF down conversion as follows:

$$s_l(mT_s) = \sqrt{\rho_l} b_l(mT_s - \tau_l) c_l(mT_s - \tau_l) e^{j\theta_l} + \eta(mT_s) \quad (1)$$

where $m$ is the discrete sample index, $T_s$ is the sampling period, $\rho_l$ is the received channel power, $b_l$ is the modulated bit, $c_l$ is the C/A spreading code, $\eta(mT_s)$ is the complex-valued AWGN random process with variance $\sigma_{FE}^2$, and $\tau_l$ and $\theta_l$ are the code and carrier phase parameters, respectively, which are in general time-varying. Residual frequencies components such as intermediate frequency and Doppler effects have been omitted for simplicity. The receiver generates local *carrier-phase* and *code-phase* replicas:

$$\ell_l(mT_s, \hat{\tau}_l) = c_l(mT_s - \hat{\tau}_l) e^{j\hat{\theta}_l} \quad (2)$$

where $\hat{\tau}_l$ and $\hat{\theta}_l$ are the estimated parameters for the $l$-th synchronized channel. The complex-valued accumulation product for the $k$-th coherent integration for a correlator is then:

$$x_{l,k} = x_l(kN_c T_s) = \frac{1}{N_c} \sum_{m=kN_c}^{(k+1)N_c - 1} s_l(mT_s) \ell_l(mT_s, \tau_l)^* \quad (3)$$

where $N_c = f_s T$ is the number of samples of the coherent integration period $T$, $f_s = 1/T_s$ is the sampling frequency, $(\cdot)^*$ is the complex conjugate operator, and the $k$-th coherent integration length is $kN_c T_s$, $k \in \{0,1,...\}$.

Considering multiple shifted code replicas (or correlator taps) in each channel based on the receiver hardware configuration, a post-correlation model for the $l$-th channel and the $k$-th coherent integration can be written as a function of an (arbitrary) discrete lag $\tau_i$ for the $i$-th correlator tap:

$$y_{l,k}(\tau_i) = \sqrt{\rho_{l,k}} R(\Delta \tau_i) e^{j\Delta\theta_{l,k}} + \eta_{l,k} \quad (4)$$

where $R(\cdot)$ is the autocorrelation function depicted as a triangle or peak [1], $\Delta \tau_i = \tau_{l,k} - \tau_i$, $\Delta \theta_{l,k} = \theta_{l,k} - \hat{\theta}_{l,k}$, and $\eta_{l,k}$ is the coherent accumulation of residual cross-correlation terms and AWGN. We define the discrete lag as $\tau_i = \hat{\tau}_{l,k} - \delta_i$, where $\hat{\tau}$ is the estimated *code-phase* value, $\delta_i = (i-1)d - \delta_{E-L}/2$, $i \in \{1,\dots,n\}$, is a code delay where $d$ is the correlator spacing in chips, $\delta_{E-L}$ is defined by the spacing between the earliest and latest correlators, $\delta_{E-L} \geq d$, and $n = \delta_{E-L}/d + 1$ is a fixed number of correlators in the receiver. As an example, a typical EPL tracking loop system uses $\delta_{E-L} = 1.0$, $d = 0.5$, and $n = 3$; a narrow correlator uses $\delta_{E-L} = 0.1$, $d = 0.05$, and $n = 3$ [35]. Additionally, the modulated bit has been omitted in (4) for simplicity.

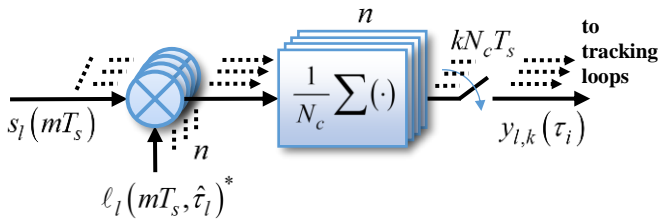

Fig. 2. A conventional GPS correlator for a single channel.

Fig. 2 shows a conventional GPS tracking loop. For a comprehensive set, Fig. 2 can be expanded to in-phase and quadrature components of the complex-valued signals, namely $s_l^I(mT_s)$ and $s_l^Q(mT_s)$, as well as for the $n$ phase-shifted correlators; otherwise, signals are considered complex-valued [7].

### B. Spoofer description

Knowing the exact position of the target receiver antenna and/or having physical access to it (e.g. PMUs) allows intermediate spoofers to carry a so-called *coherent superposition* attack [13]. It consists of synthesizing and conveying a GPS-like signal to replicate authentic *carrier-phase*, *code-phase*, and data bits, to centimeter-level accuracy for each visible open-access channel. Afterwards, the spoofer gradually increases its power so that the receiver locks to a fake correlation peak. Finally, the spoofer deliberately drags-off the correlation peak to perpetrate a PVT deviation, while maintaining lock during the attack. The reader is directed to [5] for a detailed and visual depiction of this well-known attack.

Without the loss of generality and from this point onward, we omit channel index $l$, and coherent integration instance $k$. Then the post-correlation model for a single channel and integration instance under a spoofing attack now includes additional terms:

$$y(\tau_i) = y_A(\tau_i) + y_S(\tau_i) + \tilde{\eta}$$
$$= \sqrt{\rho_A} R(\Delta \tau_i) e^{j \Delta \theta}$$
$$+ \sqrt{\rho_S} R(\tau_S - \tau_i) e^{j(\theta_S - \hat{\theta})} \quad (5)$$
$$+ \tilde{\eta}$$

where $\rho_A$ and $\rho_S$ are the authentic and spoofer powers, respectively; $\tau_S$ and $\theta_S$ are the spoofer signal *code-phase* and *carrier-phase*, respectively, and $\tilde{\eta}$ now includes additional cross-correlation terms from the spoofer. An important assumption on the spoofer model for this study is a so-called *frequency locked* attack [36], where both the authentic and spoofer are presumed to have same residual Doppler frequency during the attack, and thus is neglected in (5) and onwards. Otherwise, a sinusoid fluctuation on the spoofer peak would be observed for different $k$ integrations that could either increase, degrade, or not affect the authentic peak. The magnitude of said post-correlation output is depicted in Fig. 3 as two superimposed triangle shapes or *correlation peaks* with aligned phases. The blue triangle describes the authentic peak resulting of a typical correlator output from a tracking loop system. The more correlator taps are used, the finer resolution is seen in this triangle-shaped output.

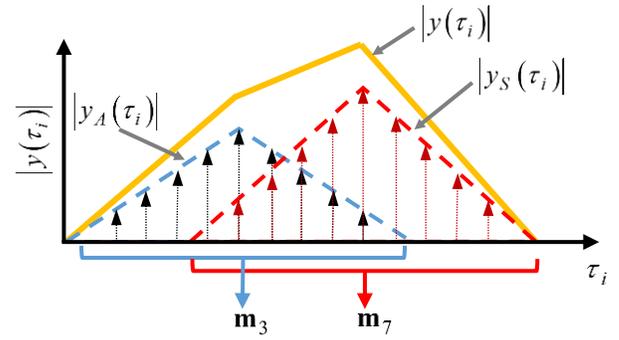

Fig. 3. A superposition of authentic and spoofed correlation triangles.

## III. DICTIONARY CONSTRUCTION AND LASSO-BASED AUTHENTICATION

We begin the problem formulation by assuming real-valued terms initially, and expanding to a comprehensive complex domain afterwards. Assuming a two-stage correlation process (before tracking loops) where carrier wipe-off occurs first, and code sample-wise multiplication and integration follows, we postulate a bank of local codes typically stored in the receiver's non-volatile memory. In the following, we express such bank of replicas in a matrix form using $n$ discrete replicas with consecutive *code-phases*:

$$\mathbf{C} = [\mathbf{c}_1, \dots, \mathbf{c}_i, \dots, \mathbf{c}_n]^T \quad (6)$$

where $\mathbf{C} \in \mathbb{R}^{n \times N_c}$, $\mathbf{c}_i = c(mT_s - \tau_i), m \in \{1,\dots,N_c\}$ is the $i$-th single-period shifted local code replica in column-vector format; and $\tau_i = \hat{\tau} - \delta_i$. This set of replicas will be used to assess the alignment of individual received signals.

Similarly, we define a *high-resolution* set of normalized and noiseless signals with $p$ discrete *code-phases*, and disregarded Doppler effects:

$$\mathbf{S} = [\mathbf{s}_1, \dots, \mathbf{s}_j, \dots, \mathbf{s}_p] \quad (7)$$

where $\mathbf{S} \in \mathbb{R}^{N_c \times p}$, and $\mathbf{s}_j = c(mT_s - \tau_j), m \in \{1,\dots,N_c\}$ is also a single-period local code replica, in column-vector format; and $\tau_j = \hat{\tau} - \gamma_j$ is the signal delay. The term *high-resolution* develops from a finer-granularity of *code-phases* between consecutive $\mathbf{s}_j$ signals. The signals $\mathbf{s}_j$ are introduced to represent ideal received signals of various delays. The received signal delays might not exactly match the set of discrete delays represented by $\mathbf{c}_i$ due to channel-induced random delays, which requires additional attention. Thus the higher resolution *code-phases* are defined for the received signals by the delay $\gamma_j = \left(j-1-\lfloor F_p/2 \rfloor\right)d/F_p - \delta_{E-L}/2$, $j \in \{1,\ldots,p\}$, along with a finer signal spacing $d_p = d/F_p$. For both *code-phase* and signal spacing, $F_p$ is called *p-factor* and defines the *high-resolution* factor between $n$ correlator taps and $p$ shifted code signals, i.e., $p = nF_p$. In particular, $p = n$ for $F_p = 1$ will correspond to the same delay grid of both received and replica signals.

Finally, we define a normalized real-valued *dictionary* of triangle replicas by correlating $p$ high-resolution code shifted signals with a bank of $n$ replicas:

$$\mathbf{M} = \mathbf{CS} = \left[\mathbf{m}_1, \ldots, \mathbf{m}_j, \ldots, \mathbf{m}_p\right] \quad (8)$$

where $\mathbf{M} \in \mathbb{R}^{n \times p}$, is the dictionary of correlations of ideal received signals (with $p$ possible code-phases) with local replica signals (with $n$ possible code-phases). In other words, the code-phase grid of the received signals is $F_p$ times finer than the code-phase grid of replicas. Here, $\mathbf{m}_j = \mathbf{Cs}_j$, is a triangle shape correlation output of a single-period local code signal, with delay $\tau_j$, with the bank of local replicas $\mathbf{C}$. Fig. 4 shows a visual representation of matrix $\mathbf{M}$ of $n$ correlation taps and $p$ shifted triangles.

With the defined dictionary matrix, the post-correlation signal can be modeled as follows:

$$\begin{pmatrix} y_1 \\ \vdots \\ y_n \end{pmatrix} = \underbrace{\begin{pmatrix} m_{1,1} & \cdots & m_{1,p} \\ \vdots & \cdots & \vdots \\ m_{n,1} & \cdots & m_{n,p} \end{pmatrix}}_{\mathbf{M}} \underbrace{\begin{pmatrix} \beta_1 \\ \vdots \\ \beta_p \end{pmatrix}}_{\boldsymbol{\beta}} + \eta \quad (9)$$

where $\mathbf{y} \in \mathbb{R}^{n \times 1}$ is the received $l$-th channel, $k$-th coherent integration (omitted) post-correlation model after carrier wipe-off, $y_i = y(\tau_i) = \sqrt{\rho}R(\Delta\tau_i)\cos\Delta\theta + \eta, i \in \{1,\ldots,n\}$ is the $i$-th correlation tap output; $m_{i,j} \in \mathbb{R}^{n \times p}$ is the $i$-th correlation tap for the $j$-th signal shift, i.e., $m_{i,j} = R(\tau_j - \tau_i)$, and $\boldsymbol{\beta} \in \mathbb{R}^{p \times 1}$, i.e., $\beta_j = \beta(\tau_j)$, is a sparse column-index selector.

In a normal operation of the receiver, the sparse vector $\boldsymbol{\beta}$ should select one triangle replica (column) from the dictionary $\mathbf{M}$ that best assimilates the *code-phase* of the received signal triangle $\mathbf{y}$. The optimal $\boldsymbol{\beta}$ can thus be recovered through solving the following LASSO minimization problem:

$$\hat{\boldsymbol{\beta}} = \underset{\boldsymbol{\beta}}{\operatorname{argmin}} \left\{ \frac{1}{2} \|\mathbf{y} - \mathbf{M}\boldsymbol{\beta}\|_2^2 + \lambda \|\boldsymbol{\beta}\|_1 \right\} \quad (10)$$

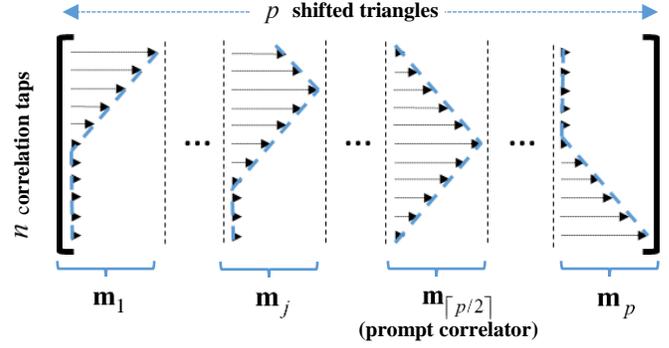

Fig. 4. Dictionary matrix of correlation triangle replicas.

where $\lambda$ is a tuning parameter that controls the amount of regularization of the sparse solution [32]. The first component in the objective function in (10) attempts to select columns of the dictionary matrix to match the received signal, while the second term encourages a sparse solution. In a successful detection of a spoofer attack, two non-zero entries in the selector $\hat{\boldsymbol{\beta}}$ are expected, e.g, $\hat{\beta}_3$ and $\hat{\beta}_7$ (see Fig. 3 for reference). It is worth noting that (10) can be reformulated into a small-to-moderate sized convex quadratic program, which can be efficiently and reliably solved. Additionally, norm-2 regularizations were explored in (10), but norm-1 showed superior robustness because it promotes sparsity.

*A. In-phase and quadrature LASSO*

In a more comprehensive problem formulation and similar to common GPS receiver tracking loops, we approach the case for $\mathbf{y} \in \mathbb{C}^{n \times 1}$ to account for spoofer peak *carrier-phase* rotations and complex-valued AWGN. We split the received post-correlation vector into its in-phase and quadrature components:

$$\begin{aligned} \mathbf{y} &= \mathbf{y}^I + i\mathbf{y}^Q \\ y_i^I &= y^I(\tau_i) = \sqrt{\rho}R(\Delta\tau_i)\cos\Delta\theta + \eta^I \\ y_i^Q &= y^Q(\tau_i) = \sqrt{\rho}R(\Delta\tau_i)\sin\Delta\theta + \eta^Q \end{aligned} \quad (11)$$

Similarly, we split the selector output, i.e., $\boldsymbol{\beta} = \boldsymbol{\beta}^I + i\boldsymbol{\beta}^Q$. We then expand the objective function in (10) to solve for both in-phase and quadrature components, either jointly or separately:

$$(\hat{\boldsymbol{\beta}}^I, \hat{\boldsymbol{\beta}}^Q) = \underset{\boldsymbol{\beta}^I, \boldsymbol{\beta}^Q}{\operatorname{argmin}} \left\{ \begin{aligned} &\frac{1}{2}\|\mathbf{y}^I - \mathbf{M}\boldsymbol{\beta}^I\|_2^2 + \lambda\|\boldsymbol{\beta}^I\| \\ &+\frac{1}{2}\|\mathbf{y}^Q - \mathbf{M}\boldsymbol{\beta}^Q\|_2^2 + \lambda\|\boldsymbol{\beta}^Q\| \end{aligned} \right\} \quad (12)$$

Finally, we obtain the magnitude of both in-phase and quadrature outputs:

$$|\hat{\boldsymbol{\beta}}| = |\hat{\boldsymbol{\beta}}^I + j\hat{\boldsymbol{\beta}}^Q| \quad (13)$$

From this point onwards, the next sections will assess the proposed spoofer detection method by using the magnitude of $\hat{\boldsymbol{\beta}}$, as in (13).

## IV. THE MULTI-LASSO TECHNIQUE

For a fixed set of correlator taps in a GPS receiver, the proposed method is able to detect peaks in a discrete grid. This restriction occurs when $F_p = 1$, and thus $p = n$. For example, if $\delta_{E-L} = 1.0$, $d = 0.1$, and $F_p = 1$, the discrete grid for detection is:

$$\delta = [-0.5, -0.4, \ldots, 0.0, 0.1, \ldots, 0.5]^T. \quad (14)$$

If a detected peak's *code-phase* does not fall on this grid, e.g., at 0.04, a *peak-splitting* occurrence is observed, as the peak falls between correlator taps 0.0 and 0.1. This can cause energy being split between two coefficients in $\hat{\boldsymbol{\beta}}$ and potentially incite a miss-detection based on a threshold level.

Based on this motivation, we attempt to overcome said *peak-splitting* phenomena by increasing the grid resolution by setting $F_p > 1$ and $p > n$. As mentioned in Section III, the high-resolution *p-factor* defines a finer signal spacing in the p-domain of the dictionary matrix, as opposed to a fixed bank of $n$ replicas coming from the configuration of the receiver. Overall, the *p-factor* increases the number of possible shifted triangle columns. These shifted triangles are correlation combinations of code replicas and signals, with *code-phases* $\delta_i, i \in \{1, \ldots, n\}$ and $\gamma_j, j \in \{1, \ldots, p\}$, respectively. Additionally, the $n$ correlators require no modification in the receiver, i.e., it can be seen as an artificial increase in grid resolution. To achieve this, we propose a method for the $p > n$ case to match the artificially generated high-resolution shifts to $n$ receiver correlator taps. Following the example in (14), setting $F_p = 5$ artificially increases the resolution from $d = 0.1$ to $d_p = 0.02$. Now, additional peak *code-phases* of $[-0.04, -0.02, 0.0, 0.02, 0.04]$ are found on the detection grid around the prompt correlator tap 0.0.

We begin the *multi-LASSO* formulation by generating a single high-resolution dictionary matrix $\mathbf{M}$ by setting $F_p > 1$ and $p > n$. We then proceed to split said fat matrix into $F_p$ individual square $n \times n$ matrices; this, to match the $n$ correlator taps of the receiver. Each decimated matrix is built by de-interleaving the columns of the original fat matrix as follows (where MATLAB notation is used):

$$\mathbf{M}_K = \mathbf{m}(K : F_p : end), K \in \{1, \ldots, F_p\}. \quad (15)$$

For example, a dictionary matrix with $\delta_{E-L} = 1.0$, $d = 0.1$, $n = 11$, and $F_p = 5$, has size $11 \times 55$. We build five individual $11 \times 11$ matrices from said matrix by taking columns $[1,6,11,\ldots,51]$ for $\mathbf{M}_1$, $[2,7,12,\ldots,52]$ for $\mathbf{M}_2$, etc. Each individual dictionary matrix can be seen as a delayed version of a square matrix for $F_p = 1$ and $n = p$, delayed by $d_p$. We then implement a *multi-LASSO* technique by adjusting (12) to include each $\mathbf{M}_K$ jointly in the $\ell 1$-minimization function as follows:

$$(\hat{\boldsymbol{\beta}}^I_{K=1\ldots F_p}, \hat{\boldsymbol{\beta}}^Q_{K=1\ldots F_p}) =$$

$$\underset{\boldsymbol{\beta}^I_{K=1\ldots F_p}, \boldsymbol{\beta}^Q_{K=1\ldots F_p}}{\operatorname{argmin}} \left\{ \begin{array}{l} \frac{1}{2} \sum_{K=1}^{F_p} \left\| \mathbf{y}^I - \mathbf{M}_K \boldsymbol{\beta}^I_K \right\|_2^2 + \sum_{K=1}^{F_p} \lambda_K \left\| \boldsymbol{\beta}^I_K \right\|_1 \\ + \frac{1}{2} \sum_{K=1}^{F_p} \left\| \mathbf{y}^Q - \mathbf{M}_K \boldsymbol{\beta}^Q_K \right\|_2^2 + \sum_{K=1}^{F_p} \lambda_K \left\| \boldsymbol{\beta}^Q_K \right\|_1 \end{array} \right\}$$

(16)

Similar to (13), we combine each in-phase and quadrature outputs to obtain $\hat{\boldsymbol{\beta}}_K, K \in \{1, \ldots, F_p\}$ magnitudes. Moreover, since each vector is of size $n$, their entries can be directly matched to the receiver correlator taps. Specifically, we choose the maximum output among all $\hat{\boldsymbol{\beta}}_K$ outputs for the $i$-th correlator tap:

$$\hat{\beta}_{i,\max} = \underset{\hat{\beta}_{K=1\ldots F_p, i}}{\arg\max} \left\{ \hat{\beta}_{1,i}, \ldots, \hat{\beta}_{K,i}, \ldots, \hat{\beta}_{F_p, i} \right\} \quad (17)$$

After finding the maximum peak for taps $i \in \{1, \ldots, n\}$ from all $\hat{\boldsymbol{\beta}}_K$ vectors, we obtain $\hat{\boldsymbol{\beta}}_{\max} \in \mathbb{R}^{n \times 1}$. The optimization technique deals with individual square matrices of size $n \times n$ per LASSO computation, thus making the solution numerically more robust. Additionally, the objective function can be computed individually and not necessarily jointly, but our simulations show that joint computation is faster.

### A. Peak-sensitivity response

To assess the sensitivity of our optimization technique for different configurations, that is, different *p-factors*, we utilize a similar concept of the impulse response in a low-pass filter; we name it *peak-sensitivity response* (PSR). First, we generate a synthetic signal with authentic and spoofer peaks as the input:

$$\mathbf{y}_{Aj, Sj} = \mathbf{C} \left( \sqrt{\rho_A} \mathbf{s}_{Aj} + \sqrt{\rho_S} \mathbf{s}_{Sj} + \sqrt{\rho_\eta} \boldsymbol{\eta} \right) \quad (18)$$

where $Aj$ and $Sj$ for $j \in \{1, \ldots, p\}$ are indices corresponding to the authentic and spoofed signals selected from matrix $\mathbf{S}$ in (7), $\rho_A$ and $\rho_S$ are the respective power levels, and $\boldsymbol{\eta}$ is complex-valued noise with power level $\rho_\eta$. Additionally, the spoofer phase is neglected. We use (18) as input to the proposed method and we evaluate a single correlator tap, e.g., $\hat{\beta}(0.3)$ for *code-phase* 0.3, as the output response. We "swing" a spoofer peak with fixed nominal carrier-to-noise ratio (CNR) through a high-granularity grid, i.e., $\mathbf{s}_{Sj}, j \in \{1, \ldots, p\}$, as a stimulus (or impulse) to generate the PSR plot for said correlator tap index. Next, we fix the authentic peak at the center, i.e., $\mathbf{s}_{Aj}, j = \lceil p/2 \rceil$. We assess the PSR of our system with the following configuration: $\delta_{E-L} = 1.0$, $d = 0.1$, $n = 11$, $f_s$ at 25 MHz, and a spoofer peak relative-to-authentic power of 0 dB. This implies a power-matching scenario. A granularity of $d_p = 0.01$ chip is used for the grid of *code-phases* and a strong signal with CNR of 50 dB-Hz and 20 msec coherent integration length is simulated to test sensitivity in nominal conditions [1].

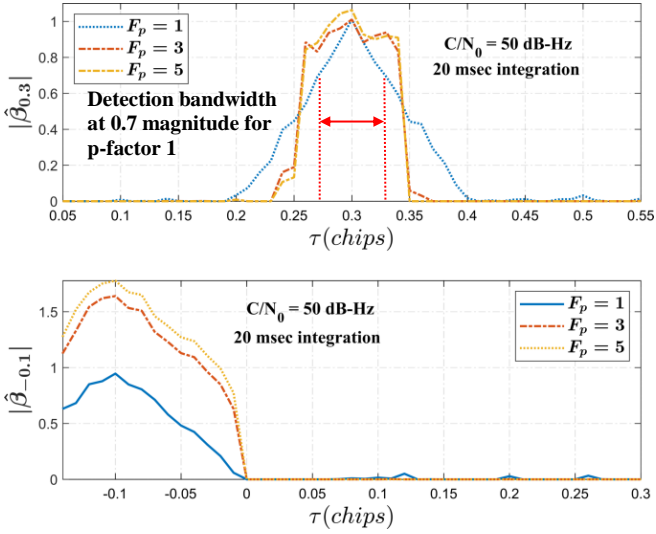

Fig. 5. PSR plot for proposed technique for spoofer peak sensitivity with *p-factor* of 1, 3, and 5, for correlator tap 0.3 (top), and correlator tap -0.1 (bottom).

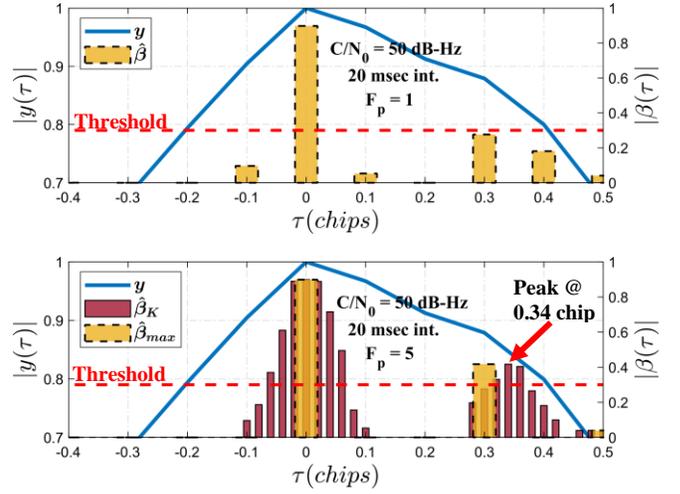

Fig. 6. Normalized received post-correlation vector **y** with simulated *code-phase* of 0.34, and CNR of 50 dB-Hz. Proposed method output with *p-factor* of 1 (top), vs *multi-LASSO* output with *p-factor* of 5 (bottom).

Fig. 5 shows a PSR evaluation for $F_p = 1$, $F_p = 3$, and $F_p = 5$ with the proposed technique in (12), and *multi-LASSO* technique in (16), for the correlator tap 0.3 (top) and tap -0.1 (bottom). The y-axis corresponds to the magnitude of the output as in (13), and the x-axis is the simulated spoofer delay, $\tau_j$. Similar to a discrete Fourier transform analysis, we evaluate a *detection bandwidth* from the PSR plot by observing the output of the optimization technique at the fixed correlator tap, i.e., $\hat{\beta}(0.3)$. On the top curve, the observed *detection bandwidth* for $F_p = 1$ at magnitude 0.7 is approximately 0.05. This translates to a sensitivity gap between neighbor correlator taps. On the other hand, a steeper curve and improved *detection bandwidth* is observed for $F_p > 1$ methods. This bandwidth corresponds to the size of the correlator spacing $d$ and translates to almost non-existent sensitivity gaps due to the increased granularity. Similar to an impulse response through a low-pass filter, the *detection bandwidth* becomes flatter and the PSR response achieves a steeper slope (roll-off factor) for $F_p > 1$.

Similarly, the bottom graph of Fig. 5 shows the PSR at $\hat{\beta}(-0.1)$, where the observed correlator tap is next to the authentic peak tap at 0.0. For the case of $F_p = 5$, a sensitivity of up to -0.02 *code-phase* at a magnitude of 0.7 is discernible from the *detection bandwidth* before reaching the 0.0 tap. This translates to a potential decomposition of the spoofer peak as close as 0.02 chips to the authentic peak, with this configuration. Additionally, a magnitude increase is observed at the output of $\hat{\beta}(-0.1)$ (bottom of Fig. 5) from the LASSO numerical outputs. This is due to a potential energy absorption between both peaks and actually aids in the sensitivity for $F_p > 1$ near the prompt tap.

## V. SIMULATIONS AND RESULTS

In this section, we perform a comprehensive set of tests that verify the proposed model in (12) and (16) for detection of a spoofer attack in the received post-correlation vector **y**. We test the proposed model on two different scenarios: synthetic generated GPS-like signals, and a real dataset. For both scenarios, we assess the selector output for two dictionary matrices: (1) $p = n$, for *single-LASSO*; and (2) $p > n$, for *multi-LASSO*.

We evaluate our optimization technique using the MATLAB-based convex-optimization solver CVX [37] along with synthetic data. We use standard cross-validation (CV) methods to tune the parameter $\lambda$ for the simulations.

First, a synthetic simulation is presented to demonstrate the advantage of using a *p*-factor greater than one. Then, a series of Monte Carlo simulations are run to characterize the DER in various scenarios. Additionally, we assess the effects of coherent integration length for enhanced CNR. To discern between noisy peaks and the authentic and spoofer peak locations, we run simulations to evaluate the PFA. Finally, the developed model is tested on data from the Texas Spoofing Test Battery (TEXBAT) [36].

### A. A synthetic simulation for multi-LASSO

We begin the evaluation with a visual instance. Fig. 6 shows a synthetic simulation of a received signal by using (18). We simulate an authentic peak with *code-phase* of 0.0, and a spoofer peak at -3 dB relative-to-the-authentic power with a *code-phase* of 0.34. The correlator parameters used are $\delta_{E-L} = 1.0$, $d = 0.1$, $n = 11$. We evaluate for both $F_p = 1$ and $F_p = 5$. A sampling rate of 25 MHz and a CNR of 50 dB-Hz is used. We have chosen this CNR as a nominal value measured in a well-known real dataset from TEXBAT [36]. On the left y-axis, the synthetic received signal post-correlation **y** is seen as the blue triangle, and on the right y-axis we have $\hat{\beta}$ outputs (please note y-axis ranges). The x-axis shows the correlator tap outputs. The top graph of Fig. 6 shows a split of 0.3 and 0.4 chips near the spoofer *code-phase* location due to its coarse grid. The transversal dotted red line shows a threshold level of 30%, or -10.5 dB of the normalized authentic signal power. This means that a spoofed-over-authentic peak power level at less than -10.5 dB will remain undetected. The threshold level is calculated from the normalized power. The threshold will be further discussed in Section V-C. Furthermore, this result shows the *peak-splitting* phenomenon discussed in Section IV. Due to this, the peak detection at phase 0.3 for $F_p = 1$ (top) is under the threshold line. The bottom graph of Fig. 6 shows the case for $F_p = 5$ with the *multi-LASSO* technique, where the interleaved magnitude outputs of $\hat{\beta}_K$ are plotted with red bars, along with the maximized output $\hat{\beta}_{\max}$ in yellow. Due to the higher-resolution in the grid, the simulated code-phase of 0.34 is clearly detected and afterwards translated to the correlator tap of phase 0.3, where this peak is now above the threshold level, at a value of 0.42 relative to the receiver power.

### B. Monte-Carlo simulation setup

We assess the model by generating synthetic complex-valued GPS-like signals with AWGN. We use Monte-Carlo simulations for a fixed CNR level assimilating nominal GPS conditions as in [36]. Our technique is evaluated on *frequency-locked* spoofing attacks, thus the carrier frequency for both authentic and spoofed peaks is neglected. Similarly, the spoofer phase $\theta_S$ is neglected. Table 1 lists the simulation parameters for signal generation, correlators' configuration, dictionary matrix sizes, and proposed method for the next results.

A DER metric is used to account for detected peaks in the simulations. In terms of detection, the two peaks with the highest values in the sparse vector output are selected as peak candidates, i.e., authentic and spoofed. Authentic and spoofer peaks at known delays are generated as in (18). If the proposed method is unable to detect the spoofer peak at the same delay, it is considered a detection error. For power levels, three levels are used in terms of the spoofed-over-authentic signal power, in dB. For simulation scenarios, a *worst-case scenario* would be an authentic-over-spoofer signal power of 6 dB, and one msec integration length, where the spoofer is the lowest in power, thus hardest to detect with low CNR levels. For threshold level, a conservative 30% obtained heuristically is used. More details on the threshold are included in Section IV-C.

### C. Simulation results

Fig. 7 depicts the DER vs *code-phase* $\tau_i$. For this result, we simulate a spoofer delay ranging from 0.1 to 1.0 chips in a granularity of $d_p = 0.1$ chips and run 1000 Monte-Carlo simulations per delay, while the authentic peak is always at 0.0 chips. We use an integration length of 1-msec to highlight the gains of *multi-LASSO* over *single-LASSO* ($F_p = 1$). For the *worst-case scenario* of authentic-over-spoofer power of 6 dB, the *multi-LASSO* $F_p = 5$ technique is able to maintain a DER of 7.7% averaged over all taps, when compared to an average DER of 16.1% for $F_p = 1$. This shows more than two-fold gain.

TABLE I
SYNTHETIC SIMULATION CONFIGURATION PARAMETERS

| Category | Parameter | Value |
|---|---|---|
| GPS signal | No. simulated peaks | 2 |
| | $f_s$ (MHz) | 25 |
| | Nominal CNR (dB) | 50 |
| | Authentic-over-spoofer power levels (dB) | 0 dB, 3 dB, 6 dB |
| | Coherent integration lengths (msec) | 1, 5, 10, 15, 20 |
| | $\theta_S$ | Neglected |
| Correlators | Frequency lock | Yes |
| | $\delta_{E-L}$ (chips) | 1.0 |
| | $d$ (chips) | 0.1 |
| Dictionary matrix | $F_p$ | 1, 5 |
| | $n$ (correlators) | 11 |
| | $p$ (*code-phase* delays) | 11, 55 |
| | $d_p$ | 0.1, 0.01 |
| Proposed technique | Threshold | 30% |
| | $\lambda$ | 0.3009 |

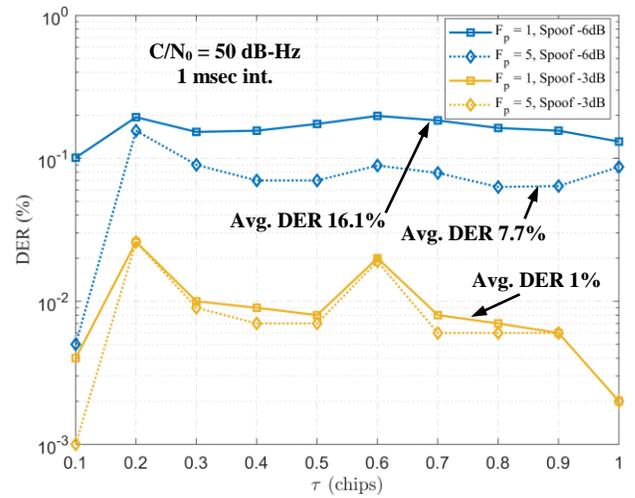

Fig. 7. Simulation results DER vs different spoofer *code-phases* $\tau_S - \hat{\tau}$ from 0.1 to 1.0 chips with CNR of 50 dB-Hz and 1 msec integration length.

For authentic-over-spoofer power of 3 dB, the two techniques see an average DER of 1% and 0.9%, for single and *multi-LASSO*, respectively. At authentic-over-spoofer power of 0 dB, the DER was essentially zero for all delays. Overall, the average DER of the proposed techniques over all spoofer power levels and discrete delays is 5.7% and 2.9%, for $F_p = 1$ and $F_p = 5$, respectively.

Similar to a BER curve, we compare the DER against different integration lengths of the received signal. The higher the integration length, the better quality of the signal as the CNR is improved with 20 periods of the 1-msec navigation bit [1]. Fig. 8 shows DER vs. coherent integration lengths of 1, 5, 10, 15, and 20 msec. For each coherent integration length and scenario, we run 1500 Monte-Carlo iterations in randomly placed spoofer peak on a grid with resolution of $d_p = 0.01$ chips. The authentic peak was placed at tap 0.0. We use the heuristically obtained threshold of 30%. The major gain can be seen for the spoofer relative power of -6 dB when $F_p = 5$ is used. An average gain of 15.7% DER is seen in such case, for all integration lengths. For the higher relative spoofer power scenario, i.e., -3 dB, 0.3% DER at 20 msec integration length is seen for both $F_p = 1$ and $F_p = 5$. Again, at authentic-over-spoofer power of 0 dB, the DER is essentially zero for all integration lengths.

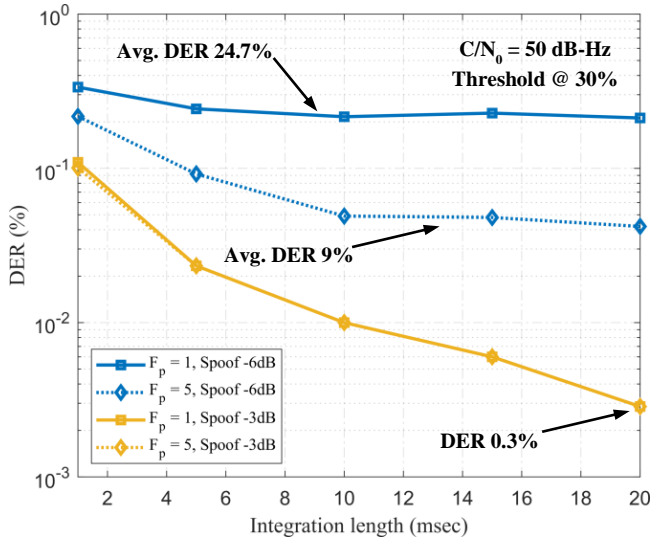

Fig. 8. Simulation results DER vs coherent integration length with CNR of 50 dB-Hz.

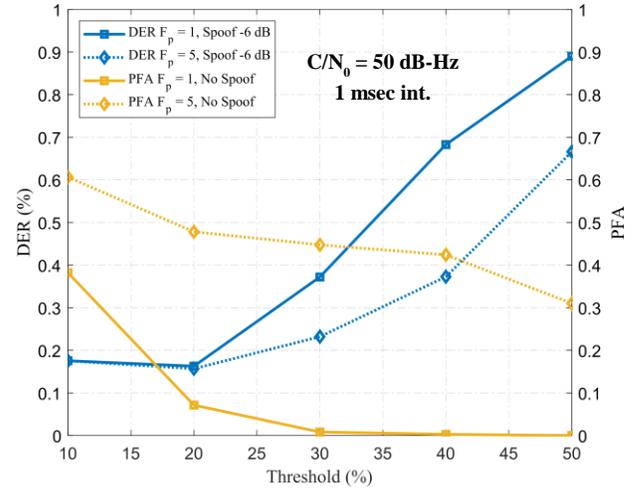

Fig. 9. Simulation results DER vs threshold vs PFA for worst-case spoofer relative power of -6 dB and 1-msec integration length.

To evaluate the impact of the heuristically obtained threshold, the PFA is assessed. A single authentic peak with nominal noise levels is simulated as we modify the threshold levels from 10 percent to 50 percent. A false alarm event is defined when a spoofer peak is wrongly detected. The worst-case scenario with an authentic-over-spoofer power of 6 dB and 1 msec integration length is assessed to estimate the PFA and DER. Fig. 9 shows results for threshold levels 10, 20, 30, 40, and 50 percent. Similarly to the results in Fig. 8, each threshold level was simulated with 1000 Monte-Carlo iterations and a randomly placed spoofer peak on a grid with resolution of $d_p = 0.01$ chips. For the PFA results, only the authentic peak centered at tap 0.0 is simulated to assess false alarm detection peaks confused with noise (that is, no spoofer is present). The *multi-LASSO* technique shows overall better DER for several threshold levels at the cost of higher PFA. This is due to a higher sensitivity for detection. We recommend that the proposed techniques be used as monitoring tools in stages as to avoid high rate of false alarms. One can use *single-LASSO* optimization for an initial detection in nominal conditions, and *multi-LASSO* can be used as a secondary stage afterwards, to detect the spoofer peak location with a higher granularity.

### D. Test with a real dataset

In this subsection, the proposed model is verified with a real dataset on a configurable SDR receiver [4], [38]. The real dataset is TEXBAT, a collection of spoofing scenarios generated at the University of Texas Radionavigation Laboratory [36]. Scenario *DS2* is selected, which represents a static example with an intermediate spoofing attack using a real-time SDR device on the target antenna [3]. The spoofer attack alters the receiver clock bias by hijacking and gradually dragging-off all channels, perpetrating their *code-phases* simultaneously. For this scenario, the final *code-phase* drag-off is around 2.1 chips, which corresponds to approximately ~610 m bias on the receiver clock. The attack begins at $t \cong 100$ s and as it drags-off, it gradually overpowers the authentic signal by 10 dB. Fig. 10 shows the attack in terms of *code-phase* difference, i.e., $\tau_S - \hat{\tau}$, for channel *PRN23*. This graph was generated using an SDR GPS receiver [38] from the Software Communications and Navigation Systems (SCNS) Laboratory at the University of Texas at San Antonio (UTSA). The spoofer peak starts dragging-off noticeably at $t = 161$ seconds by 0.2 *code-phase* (see Fig. 10 bottom for a zoomed-in version). Thus, we use snapshots of the received signal based on these attack estimates to find the spoofer peak at *code-phase* discrete values of 0.2, 0.3, 0.4, and 0.5, using the proposed algorithm.

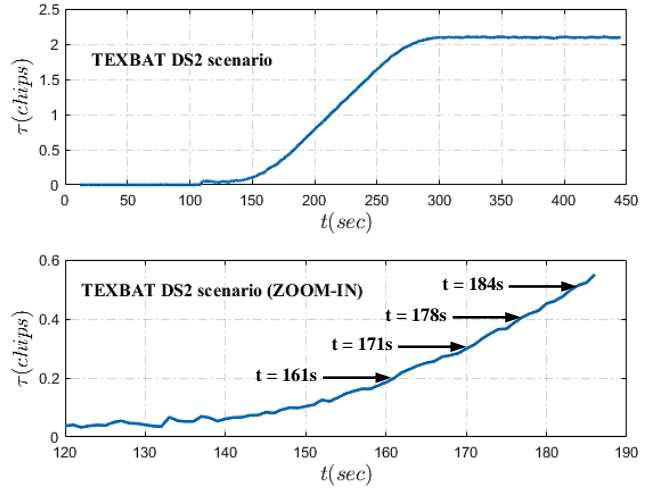

Fig. 10. TEXBAT *DS2* spoofer attack difference on *code-phase* vs authentic (top), zoomed-in around 0.5 code-phase (bottom), with markings at code-phases 0.2, 0.3, 0.4, and 0.5, respectively.

The SDR receiver for testing is able to operate in offline mode to process the dataset and extract software configurable correlator outputs [38]. TEXBAT signals were recorded with high fidelity equipment from National Instruments at 25 MHz sampling rate, and 16-bit sample resolution in interleaved in-phase and quadrature format. We configured the receiver with said parameters. Since conventional GPS receivers operate on the 1.0 chip range, we configure the correlators slightly above this chip range for visual demonstration. We configure the correlators as follows: $\delta_{E-L} = 1.6$, $d = 0.1$, and $n = 17$. We use the *multi-LASSO* technique with $F_p = 5$. We collect the correlator outputs from the SDR and apply the optimization technique on snapshots at $t = 161s$, $t = 171s$, $t = 178s$, and $t = 184s$, corresponding to estimated attack *code-phases* 0.2, 0.3, 0.4, and 0.5 chips, respectively (see Fig. 10 bottom). We also use an integration length of 1 msec. Fig. 11 depicts the results for these scenarios where the proposed technique is able to discern between both peaks at the estimated attack taps. We use a threshold of 30% as in our synthetic simulations.

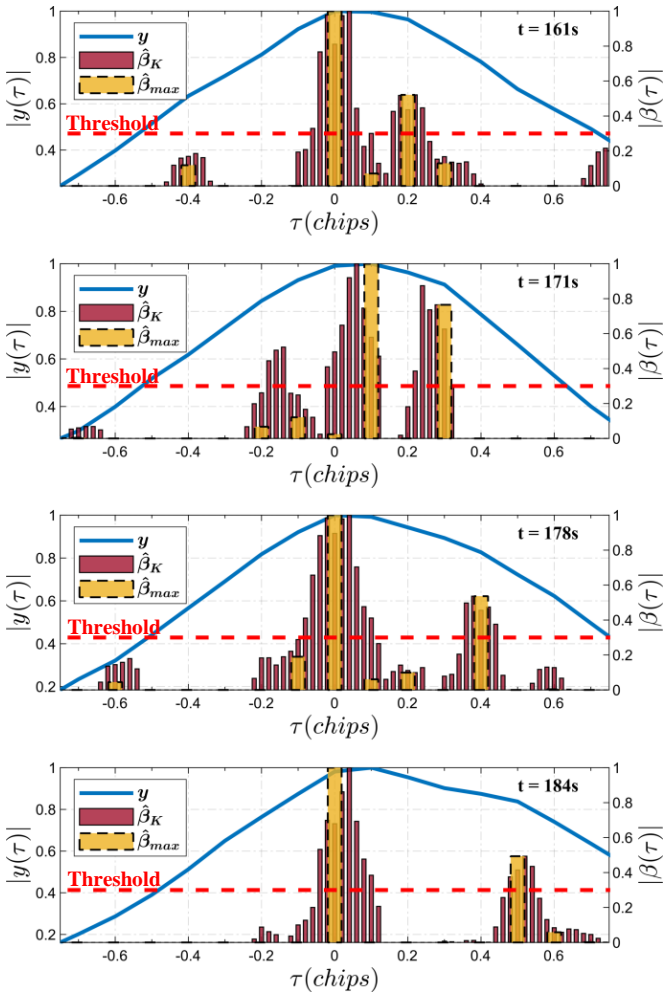

Fig. 11. *Multi-LASSO* with *p-factor* of five on TEXBAT *DS2* scenario for spoofer *code-phases* at 0.2, 0.3, 0.4, and 0.5 chips, respectively.

Working with real data introduces interesting phenomena near the vicinity of the center peak. This can be seen as a DLL discriminator *residual*, since the main peak typically shows visible red bars (see Fig. 11 at $t = 171s$). In this time slot, the main peak is found at 0.06 code-phase and is mapped to tap 0.1. Additionally, the selector might find several peak candidates near the center as multipath. Even though our proposed technique is aimed for spoofer detection, it can potentially be used as a DLL discriminator and multipath detector.

## VI. Related Work

This section provides an overview of the state-of-the-art in countermeasures against both spoofing and MP. By keeping in mind the distinction of spoofing from MP (see Section I-A), this discussion targets a qualitative distinction between spoofing and MP techniques. As discussed in Section I-C, the proposed technique falls under the advanced signal-processing category. Thus, the baseband domain of such category is considered in this section. Table 2 lists countermeasure techniques based on their baseband subcategory, countermeasure extent, whether they apply to spoofing or MP, complexity, and whether these techniques can be potentially implemented in a commercial receiver via a firmware update.

### A. Baseband domain countermeasures

Countermeasures in the *pre-correlator* subcategory rely on RF components such as the antenna, and AGC. Authors in [27] achieve blind mitigation by modifying the antenna pattern to reject low elevation angle signals where MP might appear. Authors in [39] rely on AGC power monitoring to detect spoofing over a time window. Both methods *detect* and *mitigate*, but do not *classify* (see Section I-B for definitions). Also, neither method attempts to distinguish spoofing from MP.

*Correlator* subcategory countermeasures such as the Edge or Strobe correlators adjust tap spacing to *mitigate* select MP profiles; thus, do not discern spoofing or provide a detailed MP delay profile [40]. Similarly, the multipath estimating delay lock loop (MEDLL) uses 12 correlator taps and specific metric computations to compensate for MP [28]. Signal quality monitoring techniques, namely the Vestigial Signal Defense (VSD) in [9], compute low-complexity scalar-valued metrics based on correlator peak distortions due to MP. These alarm-based methods *detect* MP per channel but do not claim any *classification* or *mitigation*. Authors in [8] also monitor correlator-based metrics and further add an observation time window to *detect* a spoofer or MP. Further, authors in [7] add hypothesis testing to such distortion metrics to further enhance *detection* between spoofer and MP. Still, these techniques do not *classify* the spoofer, e.g., provide peak delays, or MP, e.g., provide the delay profile. Authors in [10] formulate complex MP models based on certain assumptions to *classify* an MP delay profile via the maximum-likelihood estimator (MLE). Similarly, authors in [41] formulate an advanced MLE adaptive filter based on an assumed MP model. Such techniques *classify* delay profiles based on assumed MP models at the cost of high complexity. Also, such MP models are limited to specific assumptions. Finally, only MP is modeled, thus omitting spoofing attacks.

The work in [6] is more closely related to the present one. Specifically, the authors in [6] analyze correlator outputs using the Fast Fourier Transform (FFT) to *classify* peaks based on their chip delay. This method requires long non-coherent integration lengths (40 ms) and is sensitive to noise. The method in the present paper provides higher sensitivity with shorter integration lengths in similar conditions and complexity (see Section IV-A).

Finally, post-correlator techniques in [29] rely on scalar tracking loops (STL) to evaluate the code-discriminator curve and compute scalar-valued distortion metrics for MP *detection* and *rejection*. Similarly, authors in [12] use VTL outputs jointly for code-discriminator distortion metrics. These methods entail expensive receiver adaptations.

TABLE II
A STATE-OF-THE-ART COMPARISON OF BASEBAND DOMAIN ANTI-SPOOFING TECHNIQUES

| Technique | Baseband subcategory | Countermeasure extent (D/C/M) | Applies to spoofing or MP? | Complexity | Firmware update | Implementation aspects |
|---|---|---|---|---|---|---|
| Ref. [27] | Pre-correlator | M | MP | Low | No | Blind mitigation by antenna pattern tuning to avoid low elevation angle signals. |
| Ref. [39] | Pre-correlator | D | Spoofing | Low | Yes | Low-complexity power monitoring in a time observation window. |
| Ref. [40] | Correlator | M | MP | Med | No | Correlator configuration such as spacing for select MP model mitigation. |
| MEDLL [28] | Correlator | D, M | MP | Med | No | Correlators' configuration for specific MP model. Requires extra correlators and high sampling rates. |
| VSD [9] | Correlator | D | MP | Low | Yes | Distortion metrics of correlation peak. Alarm-based per channel. |
| Ref. [7] | Correlator | D, M | Both | Med | Yes | Distortion sensing of correlation peak and power monitoring. Hypothesis testing. |
| Ref. [8] | Correlator | D, C | Both | Med | Yes | Distortion sensing of correlation peak in time observation window. Hypothesis testing. |
| Ref. [10] | Correlator | D, C | MP | High | Yes | MLE based on MP model. Assumptions required. High complexity. |
| Ref. [41] | Correlator | D, C | MP | High | Yes | Advanced MLE based on non-Gaussian MP model. High complexity. |
| Ref. [6] | Correlator | D, C | MP | Med | Yes | FFT-based correlator decomposes signal into peaks. Requires long integration lengths. Noise-sensitive. |
| Proposed method* | Correlator | D, C, M[a] | Spoofing | Med | Yes | Advanced acquisition monitoring tool. Discriminates correlator peaks with high-resolution. Tunable. |
| Ref. [29] | Post-correlator | D, M | Spoofing | Med | Yes | STL discriminator-based distortions metrics. Alarm-based per channel. |
| Ref. [12] | Post-correlator | D, M | MP | High | No | VTL discriminator-based distortions metrics. Alarm-based on all channels jointly. |

[a]This method can potential implement mitigation techniques such as [31] based on smart time-based analysis of spoofer peak events.

*B. Spoofing vs. multipath*

The work in the present paper (which classifies spoofing) is similar to MP countermeasures that are able to *classify*, such as [6] and [10]. However, unlike intermittently occurring MP, intentional (smart) spoofing occurs in all GPS channels at the same time. Spoofing attacks amount to behavior change and not random interference. The work in [10] models specific MP profiles based on particular assumptions and is able to classify MP, but with high complexity. Also, because MP appears intermittently, antenna techniques such as [27] are designed to blindly reject such effects, while modified correlator techniques as in [40] only compensate for MP errors based on specific MP scenarios. Rudimentary methods such as RAIM assume a single channel is distorted per PVT epoch, and is rejected [25]; thus, it will not be able to detect an all-channel spoofing attack. The technique in [9] relies on scalar-valued metrics that *detect* potential MP distortions of the correlation peak by setting a threshold and triggering an alarm when this is surpassed.

As opposed to the previously mentioned detecting, rejecting, and compensating methods, the proposed technique offers a multi-purpose tool that *detects* a spoofing attack, and *classifies* the specific peak delays. Additionally, it is a contribution of this work to specifically model spoofing as a characteristic sparse event in the profile of peak delays, such that it can be estimated via sparse techniques. By tuning the threshold and lambda parameters, the proposed method can improve *detection* and *classification* of the attack. It is worth noting that MP is not necessarily a sparse event and its effect is less hazardous than a spoofing attack, i.e., a spoofing attack is intended to deviate the PVT solution substantially. As for *mitigation*, potential coupling of the proposed technique with auxiliary peak tracking can address this task [30], [31].

VII. CONCLUSION

In this work, a spoofing *detection* and *classification* algorithm based on LASSO is proposed to discriminate correlation peaks from a dictionary of triangle replicas. The proposed method is further extended to detect a higher-resolution grid tailored for spoofing attack delays that fall between two otherwise discrete points in the correlator tap grid. The *multi-LASSO* is able to detect spoofer peaks with a higher sensitivity without altering the receiver correlator configuration.

A peak sensitivity response method is explored to test the sensitivity of detection and define a detection bandwidth. Additionally, synthetic Monte-Carlo simulations are performed to evaluate several aspects of the proposed technique, including different integration lengths and thresholds, and relevant metrics such as DER and PFA are assessed. The proposed method is able to maintain very low DER for several scenarios and for typical receiver configurations. The proposed method achieves 0.3% DER in nominal signal-to-noise ratio (SNR) conditions for an authentic-over-spoofer power of 3 dB. Additionally, an in-house SDR receiver from UTSA is used to collect correlation points from TEXBAT, a real dataset with a spoofing lift-off attack scenario. The proposed algorithm is able to detect spoofer peaks at correlator taps 0.2, 0.3, 0.4, and 0.5 from the authentic peak, respectively.

## VIII. FUTURE WORK

In future work, real-time implementation of the proposed method is anticipated. Dynamic aspects of the implementation such as tuning of $\lambda$ based on receiver characteristics and noise levels among other aspects, are to be explored. Additionally, further examination of false alarm events and added smartness is anticipated to further enhance the proposed method. Further expansion to other GNSS signals is also proposed. Several computationally efficient algorithms for solving LASSO, such as quadratic programing (QP), the alternating direction method of multipliers (ADMM), [42], and least angle regression (LARS) [43] will be thoroughly reviewed in terms of computational requirements and compared towards possible real-time implementation. Finally, we will assess the developed signal model in terms of multipath effects and its potential applicability.

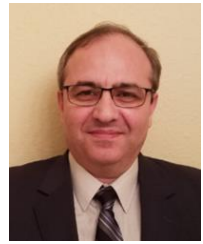
**David Akopian** (M'02-SM'04) is a Professor at the University of Texas at San Antonio (UTSA). Prior to joining UTSA, he was a Senior Research Engineer and Specialist with Nokia Corporation from 1999 to 2003. From 1993 to 1999 he was a researcher and instructor at the Tampere University of Technology, Finland, where he received his Ph.D. degree in 1997.

Dr. Akopian's current research interests include digital signal processing algorithms for communication and navigation receivers, positioning, dedicated hardware architectures and platforms for software defined radio and communication technologies for healthcare applications.

He authored and co-authored more than 30 patents and 140 publications. He is elected as a Fellow of US National Academy of Inventors in 2016. He served in organizing and program committees of many IEEE conferences and co-chairs an annual conference on Multimedia and Mobile Devices. His research has been supported by National Science Foundation, National Institutes of Health, USAF, US Navy, and Texas foundations.

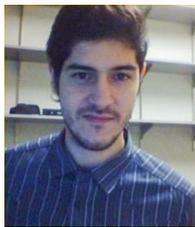
**Erick Schmidt** (S'17) received the B.S. degree (Hons.) in electronics and computer engineering from The Monterrey Institute of Technology and Higher Education, Monterrey, Mexico, in 2011, and the M.S. degree from The University of Texas at San Antonio, San Antonio, TX, USA, in 2015, where he is currently pursuing the Ph.D. degree in electrical engineering.

From 2011 to 2013, he was a Systems Engineer with Qualcomm Incorporated, San Diego, CA, USA. His current research interests include software-defined radio, indoor navigation, global navigation satellite system, spoofing mitigation algorithms, and fast-prototyping methods and accelerators for baseband communication systems.

Mr. Schmidt is a Student Member of the Institute of Navigation.

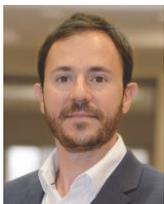
**Nikolaos Gatsis** received the Diploma degree in Electrical and Computer Engineering from the University of Patras, Greece, in 2005 with honors. He completed his graduate studies at the University of Minnesota, where he received the M.Sc. degree in Electrical Engineering in 2010, and the Ph.D. degree in Electrical Engineering with minor in Mathematics in 2012.

He is currently an Associate Professor with the Department of Electrical and Computer Engineering at the University of Texas at San Antonio. His research focuses on optimal and secure operation of smart power grids and other critical infrastructures, including water distribution networks and the Global Positioning System.

Dr. Gatsis is a recipient of the NSF CAREER award. He co-organized symposia in the area of smart grids in IEEE GlobalSIP 2015 and 2016. He served as co-guest editor for a special issue of the IEEE Journal on Selected Topics in Signal Processing on Critical Infrastructures. He was also selected to present to the 2020 NSF Engineering CAREER Proposal Writing Workshop.